\documentclass[12pt]{article}

\usepackage[]{umlaut,latexsym}
\usepackage[xdvi]{graphicx}

\usepackage{natbib}
\usepackage{amssymb}

\begin{document}

\parskip 2mm


\renewcommand{\refname}{\normalsize \bf \em References}

\title{\bf AB-INITIO STUDY OF STRUCTURE AND DYNAMICAL PROPERTIES OF
  CRYSTALLINE ICE}    

  \author{ \normalsize
W.\ A.\ ADEAGBO\footnote{
      Corresponding author. Tel.\ +49 203 379 1606;
      fax: +49 203 379 3665. 
       \newline E-mail: adeagbo@thp.uni-duisburg.de},
 A.\ ZAYAK  and P.\ ENTEL 
\\*[0.2cm]
{\small \it Institute of Physics,} \\ 
    {\small \it University of Duisburg-Essen, Duisburg campus, 47048 
Duisburg, Germany}
\\*[0.2cm]
}
\date{\small \it (Received  \today )}
\maketitle
%
\begin{abstract}
%
We investigated the structural and dynamical properties of a tetrahedrally
coordinated crystalline ice from first principles  based on density
functional theory within the generalized  gradient approximation with the
projected augmented wave method. First, we report the structural
behaviour of ice at finite temperatures based on the analysis of
radial distribution functions obtained by molecular dynamics
simulations. The results show how the ordering of the hydrogen bonding 
breaks down in the tetrahedral network of ice  with entropy increase
in agreement with the neutron diffraction data.  We also calculated
the phonon spectra of ice in  a $3\times1\times1$ supercell by using
the direct method. So far, due to the direct method used in this calculation,
the phonon spectra  is obtained without taking into account the
effect of polarization arising from dipole-dipole interactions of water 
molecules which is expected to yield the splitting
of longitudinal and transverse optic modes at the $\Gamma$-point.
The calculated longitudinal acoustic velocities from the initial
slopes of the acoustic mode is in a reasonable agreement with the neutron 
scatering data. The analysis of the vibrational density of states
shows the existence of a boson peak at low energy of translational
region  a characteristic common to amorphous systems. 
\end{abstract}
\section*{1. INTRODUCTION}
Ice, the frozen form of liquid water, is one of the most common materials on
earth and in outer space, and has important relevance to a large number
of diverse fields such as astronomy, geophysics, chemical physics, life
sciences, etc.~\citep*{cn:Victor1999}. Ice-Ih, the hexagonal form of 
ice is the most commonly known phase of ice in which
each oxygen atom has four neighbours at the corners of a 
tetrahedron. When water freezes the forces of interaction between  H$_2$O
molecules win over their thermal motion and they form a most stable
arrangement precisely with hexagonal symmetry. This is the reason why snow
flake  crystals are always hexagonal. The crystal structure of ice at
the  atomic level marks the way these crystals will look. 
The hydrogen atoms are covalently bonded to the nearest oxygen atoms
 to form water molecules which are linked to each other through
 hydrogen bonds. The high translational symmetry is not retained at
 the  level of the crystallographic unit cell. The phase diagram of
 ice is  very complex with over 12 known phases existing, (see
 Fig.~\ref{icephase}). 
 Besides the environmental importance, ice is also special because of
 the  interesting phenomena contained within its structure. The
 crystal structure of ice is very unusual because, 
while the molecules lie on a regular crystal lattice,
there is disorder in their orientations. This property leads to many
interesting characteristics in electrical polarization and conductivity.\\    
Numerous attempts have been made for the past decades to understand the nature
of the lattice vibrations of ice, in particular,
ice-Ih~\citep*{cn:Massimo1986}. Experimental information has been obtained from
infrared absorption ~\citep*{cn:Whalley1967}, Raman
scattering~\citep*{cn:Klug1978} and both coherent and incoherent inelastic
neutron scattering~\citep*{cn:Renker1969,cn:Prask1972}. On the
theoretical side, three basic approaches  have been adopted to study the
lattice modes. The earliest studies involved the application of lattice
dynamics to hypothetical proton ordered
 structures~\citep*{cn:Whalley1967,cn:Prask1972}. Later lattice
 dynamics was used to  study more realistic orientationally disordered 
structures~\citep*{cn:Nielsen1984}. The
most recent work has utilized the molecular dynamics simulation
techniques~\citep*{cn:Tse1984}. Many of the these theoretical studies involved
the use of classical modeled potential through empirical method in order to
describe the interaction of the system~\citep*{cn:Massimo1986,cn:Glen1984}. 
As result of all these works, the overall features of the lattice mode
 spectrum in the translational region (0-300 cm$^{-1}$) and in the
 librational  region (450-950 cm$^{-1}$) are reasonably well
 understood.  Potential-based empirical modelling has had some
success towards the end, but to date, there are no empirical potentials
capable of describing the ice dynamics and related properties across its
whole spectra range and describing certain key spectra features.\\    
The ab-initio method has recently gained ground not only because of its
reliability in the study of static and dynamics properties of ice
~\citep*{cn:Cote2003,cn:Morrison1999} but also because the method allow to
 model  some important features such as the ordered periodic ice
 structure~\citep*{cn:Lee1993}, and the nature of hydrogen bond in different
geometries~\citep*{cn:Xantheas1993}. Our present study  to understand the
microscopic nature and lattice vibrations of ice-Ih makes use of the  Vienna
{\it Ab Initio} Simulation Package (VASP)~\citep*{cn:Kresse96} which is
designed to perform ab-initio quantum-mechanical molecular dynamics using
pseudopotentials and a plane wave basis set and the recent implementation of
the projected augmented wave (PAW) method.\\
The computational study presented  in this work begins with the choice
 of unit cell of ice as described in Section 2, which is  followed by molecular
dynamics simulation of ice in a supercell (Section 3) with the unit
 cell replicated in all directions in order to obtain the structural
behaviour in comparison with the available experimental data. In
Section 4 and 4.1 we present the results of calculated phonon spectrum 
of ice in the translational region, which is done in a $3\times1\times1$
supercell in $y$-direction, as well as the corresponding integrated
vibrational density of states in Section 4.2.  The anomalous behaviour
of this ice structure observed in the low energy region is presented in
Section 4.3. The final Summary of this work is presented in Section 5.  
\section*{2. COMPUTATIONAL DETAILS}
As mentioned above, the calculations in this work  were carried out by
using  the Vienna {\it Ab Initio} Simulation Package (VASP)
\citep*{cn:Kresse96} which has been designed for ab-initio
quantum-mechanical molecular dynamics simulations using pseudopotentials and a
plane wave basis set and the recent implementation of projected
augmented wave (PAW) method. In order to  investigate the structural
properties of ice, we started  with the construction of ice crystal using the
Bernal and Fowler rule~\citep*{cn:Bernal1933}. 
The rule is based on a statistical model of the  position of hydrogen atoms
 produced by Pauling~\citep*{cn:Pauling1935} using the six possible
 configurations of  hydrogen atoms within Ice Ih. It is defined as
 ideal crystal based on the assumptions that:
\begin{itemize}
\item Each oxygen atom is bonded to two hydrogen atoms at a distance of 0.95 Å
 to form a water molecule;
\item Each molecule is oriented so that its two hydrogen atoms face two, of
  the four, neighbouring oxygen atoms in the tetrahedral coordination;
\item The orientation of adjacent molecules is such that only one hydrogen
  atom lies between each pair of oxygen atoms;
\item Ice Ih can exist in any of a large number of configurations, each
  corresponding to a certain distribution of hydrogen atoms with respect to
  oxygen atoms.
\end{itemize}
The schematic drawing  shown in Fig.~\ref{tetraice} satisfies one out
of the  six possible orientations of protons of the central water
molecule  according to these rules.
Each of the oxygen atoms can be linked to another oxygen by the combination of
a covalent bond plus  a hydrogen bond to form the tetrahedral arrangement of
oxygen atoms.
A unit cell of this ice was prepared in a cubic box according to
Fig.~\ref{cellice} with 8 molecules of water. All the atomic degrees
of freedom  were relaxed  using VASP with the 
projected-augmented wave (PAW) formalism  at high precision.
The optimum Monkhorst Pack $4\times4\times 4$  $k$-point was
used in addition to the generalized gradient approximation (GGA) of
Perdew-Wang in order to describe the exchange-correlation and  to give good
description of  the hydrogen bonding of water.  We used a energy cut-off
of 500 eV because the 2p valence electrons in oxygen  require a large
plane wave basis  set to span the high energy states described by the
wavefunction  close to the oxygen nucleus and also the  
hydrogen atoms  require a larger number of planes
waves in order to describe localization of their charges in real space.
The lattice constants of  the unit cell were calculated from the plot
of  the energy against the volume. The  estimated values of the lattice
 constants  are $a$ = 6.1568 {\AA}, $b$ = 6.1565 {\AA}, $c$ =
 6.0816{\AA}. The  values of $a \approx b \ne c$  imply that the
 relaxed structure is  tetragonal with ${c}/{a}$ ratio $\approx$ 0.988. The
 experimental lattice  constant reported by Blackman {\it
   et. al.}~\citep*{cn:Blackman1958} is  6.35012652 {\AA} for the
 cubic  geometry.
The final geometry obtained was used in our molecular and lattice
dynamical study of ice.
\section*{3. MOLECULAR DYNAMICS}
In our molecular dynamics simulation, the final relaxed geometry in
Fig.~\ref{tetraice} (or  Fig.~\ref{cellice} was replicated in all
directions using the calculated  values of the lattice parameters to
produce 64 molecules of water which form the hexagonal structure
shown in Fig.~\ref{ICE}. Here the molecular dynamics simulation was
done for the $\Gamma$-point only in a box of dimension 
$12.411356\times12.411356\times12.259675$ {\AA}$^{3}$ corresponding to the
density 1.01 $\mathrm{gcm^{-3}}$ to be compared to the real density of ice Ih
and ice Ic which is  0.92 $\mathrm{gcm^{-3}}$.
Molecular dynamics was carried out for 3 ps using a time step of 0.5 fs.
The angle H-O-H of the ice structure was compared with liquid water at the
different temperatures as can be seen in Fig.~\ref{angdis}. For ice structure,
the H-O-H angle was found to be 107.8$^\circ$ compared to liquid water case at
different temperature or isolated water molecule also calculated with VASP
which is found to be 104.5$^\circ$ and 105.4$^\circ$ degree,
respectively, as can be seen in Fig.~\ref{angdis}. The angle shown by
solid ice is an indication that the oxygen centre preserves its
tetrahedral structural units. Also, the range of the angular
distribution  for the high temperature water (or supercritical water)
is wider and broader than the corresponding liquid water at  ambient
temperature and the solid ice cases.\\ 
The radial distribution functions for the solid ice accumulated over a 3 ps
run are plotted at two different temperatures, 100 K and 220 K. The
$g_\mathrm{OO}$  and  $g_\mathrm{OH}$  in Fig.~\ref{ICEGR100_OO_OH}, and
 $g_\mathrm{HH}$ in Fig.~\ref{ICEGR_100_HH} at 100 K, all exhibit the
 long-range like order when compared to the short range order of liquid water.
There is a well pronounced first peak of
${g_\mathrm{OO}}$ at 2.75 {\AA} at 100 K and also the position of the first
minimum is deeper when compared to the liquid water radial distribution
functions. This result is comparable to the result obtained using the
classical TIP4P modelled potential~\citep*{cn:Igor1994}
The result of the radial distribution function of oxygen atoms,
$g_\mathrm{OO}$, oxygen-hydrogen atoms $g_\mathrm{OH}$
(Fig.~\ref{ICEGR_220_OO}),  and hydrogen atoms
(Fig.~\ref{ICEGR_220_HH}) at 220 K were fairly comparable with available
neutron diffraction scattering data of Soper
~\citep*{cn:Soper2000}. The position of the first peak (VASP calculation)
shows  very little difference from 100 K while
the height of the peak is lower for 220 K due to the effect of entropy
increase of the  hydrogen atoms  which results from re-orientation of protons
that tends to force apart more oxygen atoms to a larger distance. The
positions of the second minimum for 100 K and 220 K solid ice are,
respectively, 4.8 {\AA} and 4.9 {\AA}. The experimental result is slightly
shifted to the right to the value 5.1 {\AA} for 220 K. The deviation from
experimental value might be due to the phase of crystalline ice under
consideration.
\section*{4. LATTICE DYNAMICS}
In this study phonon dispersion curves are calculated by using the PHONON
package developed by K. Palinski~\citep*{cn:Parlinskin2002} which has
been designed to take  input data of Hellmann-Feymann forces
calculated with the help of an {\it ab-initio} electronic structure
simulation  program such as VASP.
We  carried out the lattice dynamics study of ice  by using the geometry of
eight-molecule primitive cell discussed in the Section 1.
 The calculations of force constants was carried out by considering a
$3\times1\times1$ supercell containing 24 molecules of water which is obtained
by matching 3 tetragonal unit cells. At the first step of the calculation, the
PHONON package is used to define the appropriate crystal supercell for use of
the direct method.
As done for the primitive unit cell, all the internal coordinates were
relaxed until the atomic forces were less than $10^{-4}$ eV/{\AA}. The
relaxed geometry for a $1\times1\times1$ supercell 
from the initial configurations containing 8 molecules is shown in
Fig.~\ref{cellice}. The starting geometry of the molecules in the
simulation box shown is such that no hydrogen bonds were present but the
positions of oxygen atoms follow the tetrahedral orientation. After the
relaxation, all the protons perfectly point to the right direction of oxygen
atoms and make the required hydrogen bonds necessary  as indicated  by the
dotted lines in Fig.~\ref{cellice} to preserve the tetrahedral orientation of
the ice structure. \\
Figure ~\ref{hexsym}(I) shows the Brillouin zone belonging to the relaxed
structure of our (model) ice shown in Fig.~\ref{cellice}.
Figure~\ref{hexsym}(II) shows the Brillouin zone used
in the analysis of the measured phonon spectra for the model structure of
D$_2$O ice.  Let us say once again that the relaxed structure shown in
Fig.~\ref{cellice} has the long-range orientational order while the actual
structure of ice Ih has no long-range
orientational order.  Therefore, in the analysis of the measured
phonon dispersion curves of D$_2$O, one uses another model for ice shown in
Fig.~\ref{ICE}. We have to keep this in mind  when comparing our calculated
dispersion curves with the measured frequencies.
For the evaluation of the phonon dispersion curves we have used the direct
{\it ab-initio} force constant method~\citep*{cn:Parlinski997},
whereby  the  forces are calculated with VASP via the Hellmann-Feymann 
theorem in the total energy calculations. Usually, the calculations
are done for  a supercell with periodic boundary conditions.  
In such a supercell,  a displacement ${\bf u}(0, k)$ of
a single atom induces forces ${\bf F}(lk)$ acting on all other atoms,
\begin{eqnarray}
F_\alpha(lk) = \sum_{l^\prime k^\prime \beta} \Phi_{\alpha \beta}(lk;l^\prime
k^\prime) \, u_\beta (l^\prime k^\prime).
\label{Force1}
\end{eqnarray}
This expression allows to determine the force constant matrix directly
from the calculated forces (see Parlinski {\it et. al.})
~\citep*{cn:Parlinski997,cn:Parlinskin2002}.
The phonon dispersion branches calculated by the direct method are exact for
discrete wave vectors defined by the equation
\begin{equation}
 \exp{(2\pi \imath \mathbf{k}_L \cdot \mathbf{L})}=1,
\end{equation}
where $\mathbf{L}=(L_a, L_b, L_c)$ are lattice parameters of the
supercell.
A related technique has recently been used to obtain accurate full phonon
dispersions in highly symmetric structures of $\mathrm
{Ni_2}$GaMn~\citep*{cn:Zayak2003}.\\
In order to obtain the complete information of the values of all force
constants, every atom of the primitive unit cells was displaced by 0.02 {\AA}
in both positive and negative non-coplanar, $x$, $y$ and $z$ directions to
obtain pure  harmonicity of the system.
We use a $3\times 1\times 1$ supercell which implies that 3 points in the
direction [100] are treated exactly according to the direct method. The points
are [$\zeta$00], with  $\zeta$ = 1, 1/3, 2/3.
 We calculate forces induced on all atoms of the supercell when a
single atom is displaced from its equilibrium position, to obtain the force
constant matrix, and hence the dynamical matrix. This is then followed by
diagonalization of the dynamical matrix which leads to a set of eigenvalues
for the phonon frequencies and the corresponding normal-mode
eigenvectors. The vibrational density of states (VDOS) is obtained by
integrating over  $k$-dependent phonon frequencies from the force-constant
matrix in supercells derived from the primitive molecule unit cells.\\
\section*{4.1  Phonon dispersions of crystalline ice}
 The phonon dispersion curves calculated for  our ice crystal in [100]
 direction are shown in Fig.~\ref{Cote} for low lying
 energy vibrations. According to the geometry of
 the supercell, the low-frequency (or low-energy 0-50 meV) acoustic modes can
 be  compared to Renker's inelastic neutron scattering
 measurement~\citep*{cn:Renker1973} along  the [0001] direction ($\Gamma$A) of
 hexagonal symmetry shown in Fig.~\ref{hexsym}, taken
 from  reference~\citep*{cn:Jichen1996},  though our
 calculation was done only along the Cartesian  direction [100] of the
 cubic cell. Our transverse and longitudinal acoustic (LA/TA) dispersions
 are well behaved when compared to some other modelled calculations or
 experimental results~\citep*{cn:Renker1973} in the high symmetry directions
 $\Gamma A$ of hexagonal ice as shown in  Fig.~\ref{Cote}(c).
 We can also compare our result to the Cote {\it et. al.} result
 in Fig.~\ref{Cote}(b) ~\citep*{cn:Cote2003},  where they have
 recently used the {\it ab-initio} method to obtain the phonon
 dispersions in  the translational frequency  range for the ice
 structure  in the Brillouin zone of the orthorhombic eight-molecule
 unit cell.  Altogether our LA and TA dispersions
 are better than Cote's LA/TA in comparison to the experimental curves in
 Fig.~\ref{Cote}(c). Our  dispersion curves in [100] direction can as well be
 compared to the dispersion curves obtained using a dynamical model with
 two force constants to describe the low frequencies of vibrations of hexagonal
 ice as proposed by Faure~\citep*{cn:Faure1969}.\\
Although everywhere along the  $\Gamma$-point, our  dispersions are completely
 degenerate in the optic  region whereas  Cote's dispersions in
 Fig.~\ref{Cote}(b) show some splittings, the so-called
 longitudinal/transversal optic (LO/TO) splittings whose origins is
 explained   below,  while at the zone boundary, some  of the dispersions are
 non-degenerate unlike our  results.
 Our inability to reproduce these splittings at the
 $\Gamma $-point is due to the direct method approach in which absolute
 periodicity of  the crystal according to Born-von K\'arm\'an conditions was
 considered.  The splitting of LO and TO branches for long wavelengths occurs
 in almost all crystals which are heteropolar (partially ionic such as GaAs) or
 ionic (such as NaCl) at the $\Gamma$-point, and only for infrared active
 modes~\citep*{cn:Srivastava1990}. The long-range part of the Coulomb
 interaction causes the splitting of the $k$ = 0 optic modes raising
 the  frequency of LO modes above those of TO modes. 
The long-range part of the Coulomb interaction
 corresponds to the macroscopic electric field arising from ionic
 displacements. Ice is a tetrahedrally covalently bonded polar
 system whose dipole-dipole interactions give rise to  the electric field when
 they are  disturbed. The origin of the splitting is
 therefore the  electrostatic field created by long wavelength modes of
 vibrations in such crystals. Usually a microscopic electric field influences
 only the LO modes  while TO modes remain unaltered. The field therefore
 breaks the Born-von  K\'arm\'an conditions, as a consequence with a direct
 method only finite wave  vector ${\bf k}\ne 0$ calculations are
 possible. Elongated sub-supercells are needed to recover the ${\bf
 k}\rightarrow 0$ limit of the LO phonon branch~\citep*{cn:Parlinskin2002}.\\
In our result there are two transverse acoustic branches which are highly
degenerate and a  longitudinal acoustic branch. The first optical branch of
the dispersion curves is degenerate with the transverse acoustic branches at
energy $\sim$9.0 meV.
The transverse and longitudinal velocities of sound are calculated from the
initial  slopes of the corresponding transverse and longitudinal acoustic
branches of the dispersion curves in the long wavelength limit.
The experimental  values of velocities reported in Table~\ref{velTab}
are those of longitudinal and transverse sound waves propagating along the
$c$-direction of single crystals of ice at  257 K. It is well
known that velocities of sound depend much on the direction of propagation
and also on the temperature. Inelastic X-ray scattering data from water at 5
°C shows a variation of the sound velocity from 2000 to 3200 m/s in the
momentum range of 1-4 $\mathrm{nm^{-1}}$. The so-called transition from {\it
  normal} to {\it fast} sound in liquid water at $\approx$ 4 meV, the energy
of sound excitations which is equal to the observed second weakly dispersed
mode, was reported to be due to the reminiscent of a phonon branch of ice Ih
 of known optical character~\citep*{cn:Sette1996}. We can conclude that our
calculated values of longitudinal velocity, $v_L$ is in a reasonable range of
velocity of sound in ice along  the $[100]$-direction chosen for our
calculation. We must also stress the fact that our phonon dispersions were
calculated at 0 K.

\section*{4.2  Vibrational density of states of crystalline ice}
In order to understand the mode of collective vibration of molecules of water
in ice from a spectroscopic point of view, we need to consider the three normal
modes of an isolated water molecule shown in Fig.~\ref{tmodes} as  phases of
water vapour; liquid water and ice consist of distinct H$_2$O molecules
recognized by Bernal and Fowler in 1933, which explained one of the factors
leading to the Pauli model of the crystal structure of ice. The fact that the
forces between the molecules are weak in comparison with the internal bonding
results in a simple division of the lattice modes into three groups involving
the internal vibrations, rotations, and translations of the molecules. The
frequency of the first two groups depend primarily on the mass of the hydrogen
or deuterium nuclei, and the frequencies of the translations depend on the
mass of the whole molecule~\citep*{cn:Victor1999}.
A free H$_2$O molecule has just three normal modes of vibration illustrated in
Fig.~\ref{tmodes}. The comparatively small motions of the oxygen atoms are
required to keep the centre of mass stationary, and these motions result in
the frequency $\nu_3$ being slightly higher than $\nu_1$; these depend on the
force constant for stretching the covalent O-H bond, while the bending mode
$\nu_2$ depends on the
force constant for changing the bond angle. In the vapour the free molecules
have a rich rotation-vibration infrared spectrum~\citep*{cn:Benedict1956}, from
which the frequencies of the molecular modes are deduced to be:\\
\begin{center}
$\nu_1$ = 3656.65 cm$^{-1}$ $\equiv$ 453.4 meV,\\
$\nu_2$ = 1594.59 cm$^{-1}$ $\equiv$ 197.7 meV,\\
$\nu_3$ = 3755.79 cm$^{-1}$ $\equiv$ 465.7 meV.
\end{center}
For ice the band around 400 meV is thus ratified with the O-H bond stretching
modes $\nu_1$ and $\nu_3$. The frequencies are thus lowered from those of the
free molecules by the hydrogen bonding to the neighbouring molecules, but as a
single molecule cannot vibrate independently, this coupling also leads to
complex mode structures  involving many molecules.\\
We can now discuss the vibrational density of states, VDOS, for H$_2$O
ice based on the lattice dynamics obtained from the results of our calculation
and compare them to some of the well known spectra of ice such as infrared and
Rahman spectra  and inelastic neutron scattering data.
 The total VDOS calculated from the phonon dispersions for our ice structure is
 shown in Fig.~\ref{totviblat}. Also shown in Fig.~\ref{viblat} is the
 corresponding partial VDOS for both hydrogen  and
the  oxygen atoms in the ice system. We note that these phonon DOS are not
complete since the summation is not done over the whole Brillouin zone, but
only  in the [100] direction of the cubic symmetry.
The distribution of the partial DOS is given by
\begin{eqnarray}
g_{\alpha,k}(\omega) = \frac{1}{nd\Delta \omega} \sum_{{\bf k},j}
\left.|e_\alpha(k;{\bf k},j)\right|^2 \delta_{\Delta \omega}(\omega
-\omega({\bf k},j)),
\label{pdos}
\end{eqnarray}
where $e_\alpha(k;{\bf k},j)$ is the $\alpha$-th Cartesian component of the
polarization vector for the $k$-th atom; $n$ is the number of sampling points
and $d$ is the dimension of the dynamical
matrix~\citep*{cn:Parlinskin2002}. The total VDOS is calculated by
summing all the partial contributions. Figure~\ref{totviblat}  shows
the total VDOS together with the
full phonon dispersion curves along the [100] direction. Also shown in
Fig.~\ref{totviblat2} is the enlargement of the intermolecular frequency range
on which we superimpose the inelastic neutron-scattering spectra data
extracted from  Ref.~\citep*{cn:Cote2003}. The comparison is made with
the ice Ih data though our ice structure is not perfectly hexagonal but still
there is very little difference between the neutron data for ice Ih and ice
geometry used in our calculation in the translational region as explained
below.
There are well defined separated peaks in the whole range of the
vibrations. The illustrative discussion in Fig.~\ref{tmodes} can be well
understood if we consider the partial DOS in Fig.~\ref{viblat}. The covalently
O-H stretching mode of both phase and anti-phase, analogous to the frequencies
$\nu_1$ and $\nu_3$ for isolated free water molecules, can be seen clearly in
Fig.~\ref{viblat}(b) in the energy range (350-410 meV) or frequency range
(3010-3400 cm$^{-1}$). We can notice that the collective motion of oxygen is
almost static when compared to the collective contributions from the hydrogen
atoms. According to the Rahman spectra, a strong peak is observed at 382.3 meV
(3083 cm$^{-1}$ at 95 K)~\citep*{cn:Bertie1967,cn:Whalley1967} for
${\mathrm{D_2O}}$. If we take into account the mass difference between
deuterium and hydrogen atoms (i.e., isotope effect), the peaks which are
observed at 2950, 3000 (very short), 3250, and 3270 cm$^{-1}$ are in good range
when compared to the experimentally observed values for $\nu_1$ and $\nu_3$.
In the intra-molecular bending region, analogous to the frequency $\nu_2$
for an isolated water molecule (1580-1680 cm$^{-1}$), there is an
interesting feature. Our results show that, of all the contributions resulting
from the collective motion of hydrogen atoms as contribution from collective
motion  of oxygen atoms  is recessive, only {\it one} of the components of the
collective  motions of hydrogen atoms contributes to the intra-molecular
bending modes and it is one that is dominant. Figure~\ref{viblat}(b) gives an
example of such contribution
being dominated mainly by the $y$-component of the intra-molecular vibration
of the O-H. This means that intra-molecular bending of the angular motion
takes place mostly in one direction.\\
Tanaka has identified hydrogen-bond bending modes
with  negative expansion coefficients associated with this
region~\citep*{cn:Tanaka1998}. 
If we go further down to the low frequency region such as 600-1200
cm$^{-1}$ called the molecular librational region and then to (0-400 cm$^{-1}$)
called the molecular translational region, where we estimated the sound
velocities from the corresponding phonon dispersion curve, the VDOS peaks of
 these  modes of vibration  agree very well with the experimental
observation from inelastic neutron scattering data. The general agreement of
the features in the translational optic region is good with all the three
distinct peaks present at 400, 270 and 105 cm$^{-1}$ ~\citep*{cn:Jichen1996}.
\section*{4.3  The boson peak in ice}
The high frequency (0.1-10 THz)  or energy (0.4136-41.36 meV) excitations have
been experimentally shown to have linear dispersion relations in the
mesoscopic momentum region ($\sim$ 1-10 nm$^{-1}$). 
So many amorphous materials display low temperature
anomalies in their specific heats that they are generally regarded 
as being universal properties of the glassy state. 
These anomalies are usually of two kinds. The  first
concerns observation that, while many crystals obey the
Debye law $C\propto T^3$ for temperatures less than say, 1K, glasses
with  the same chemistry frequently display the law $C\propto T$ at 
correspondingly low temperature~\citep*{cn:Phillips1996}.
This second observation is tied in with the appearance of the
ubiquitous Boson peak (BP) in inelastic neutron and Raman 
spectra~\citep*{cn:Buchenau1987}. 
The name  Boson peak  therefore refers to the fact that the
temperature dependence of its intensity scales roughly with the
Bose-Einstein  distribution. 
Dove {\it et. al.} demonstrated that the Boson peak arises largely
from a flattening of the dispersion of the transverse acoustic modes,
in a manner similar to that which occurs in a crystalline material at
its  Brillouin zone boundary~\citep*{cn:Martin1997}. 
This idea is not new, being originally  suggested
by~\citep*{cn:Leadbetter1969}, and recently developed further by
~\citep*{cn:Elliott1996,cn:Taraskin1997a,cn:Taraskin1997b}.
A standard way of  extracting the boson peak (BP) from the vibrational
spectrum is to plot $g(E)/E^2$ (as done in Fig.~\ref{Bosonp}(b)),
since in the Debye  approximation $g(E)\approx E^2$ at low energy.   
In Fig.~\ref{Bosonp}  we show the plot of
$g(E)$ and the corresponding $g(E)/E^2$ in the translational low energy range
for the ice geometry in our calculation. According to the Debye law, it is
expected that  $g(E)/E^2 $ should be constant for the whole range of energy.
This constant relation is only obtained at energies larger than 40 meV in
agreement with the experimental observation  range  for the BP as mentioned
above. There is an anomalous sharp peak at 3.5 meV which can be ascribed
to the region of low-energy {\it excess} vibrational excitation of the
so-called BP. The reason for this peak is unknown since  we  have a 
crystalline structure for our ice geometry. The peak reveals the
anomalous behaviour of hydrogen bonding in the crystal ice which shows
similarity in the  behaviour as in the results of an inelastic neutron
scattering  study of a crystalline polymorph of SiO$_2$
($\alpha$-quartz),  and a number of silicate glasses (pure silica,
SiO$_2$) with tetrahedral coordination~\citep{cn:Harris2000}. Also amorphous solids, most
supercool  liquids and  the complex  systems show this anomalous 
character~\citep*{cn:Grigera2003}.
\section*{5. SUMMARY}
Our MD simulation, carried out  through the
analysis of  radial distribution functions of the ice crystal,
shows fair agreement in the positions of peaks in comparison with
 neutron diffraction data. \\
 The phonon dispersion calculations in [100]
 direction shows a good result especially in the acoustic region in
 comparison with experiment. Our optic modes are degenerate at the
 $\Gamma$-point due to application of the direct method used in this
 calculation which does not take into account the
effect of polarization arising from dipole-dipole interactions of
water molecules which is expected to yield a significant
effect in the splitting of longitudinal and transverse optic modes at
the $\Gamma$-point. We intend to include this polarization effect in
our future work through the calculation of effective charge tensors using the
Berry phase approach to see actually if there will be splitting. 
Our calculated longitudinal acoustic velocity
agrees well with the longitudinal acoustic velocity from inelastic
neutron scattering data. The vibrational density of states reproduces
all the features  in covalently O-H stretching region, intra-molecular
bending region, molecular  librational region as well as in the
molecular translational region, when comparing our results to some
infrared spectra, Rahman spectra as well as inelastic neutron
scattering results.\\
The analysis of the vibrational density of states shows 
a boson peak,  a characteristic feature common to amorphous systems, at low
energy  of the translational region.
The anomalous sharp peak of $g(E)/E^2$ at 3.5 meV, which can be ascribed
to the region of low-energy {\it excess} vibrational excitation of the
boson peak, might be  due to the anomalous behaviour of hydrogen
bonding since we have a perfect crystal. The origin of this anomalous 
behaviour of this abnormal peak in non-crystalline solids and
supercool liquids is still subject  of scientific debate. 
\\*[0.5cm]
\noindent {\normalsize \bf \em Acknowledgments}
\\*[0.5cm]
We acknowledge the  support by the Deutsche Forschungsgemeinschaft
(Graduate College {\em ``Structure and Dynamics of Heterogeneous
Systems''}).\\
We thank Prof. Keith Ross for allowing us to use one of
their  phonon dispersions  for comparison with our results.
Also, we  acknowledge Prof. Soper for providing us the neutron diffraction data
of radial distribution functions of liquid water and ice for fitting
our {\it ab-initio} simulation  results. \\
{\small
  \bibliographystyle{phase-trans}
  \bibliography{SDH2003adeagbo}
}

\clearpage\centerline{TABLES}

\begin{list}{}{\leftmargin 2cm \labelwidth 1.5cm \labelsep 0.5cm}

\item[\bf Table~\ref{velTab}] Velocities  of sound calculated from
 the  initial slope of the phonon dispersion curves of ice in [100] 
direction compared to the experimental result.
\end{list}
\clearpage \centerline{FIGURE CAPTIONS}

\begin{figure} 
  \centering
  \resizebox{9cm}{!}{\includegraphics*{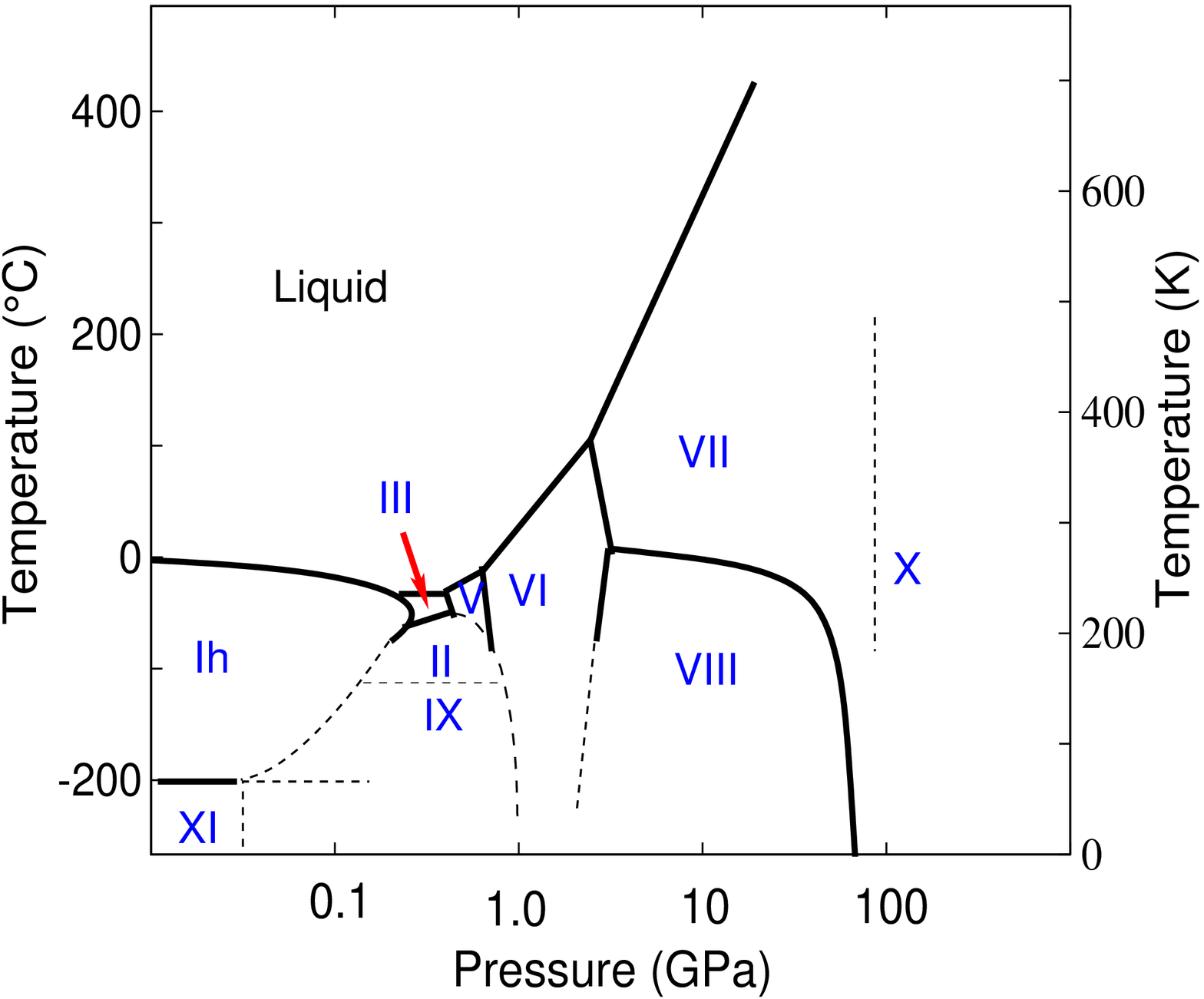}}
   \caption{ W.\ A.\ Adeagbo et. al.}
  \label{icephase}
\end{figure}

\begin{figure} 
  \centering
  \resizebox{10cm}{!}{\includegraphics*{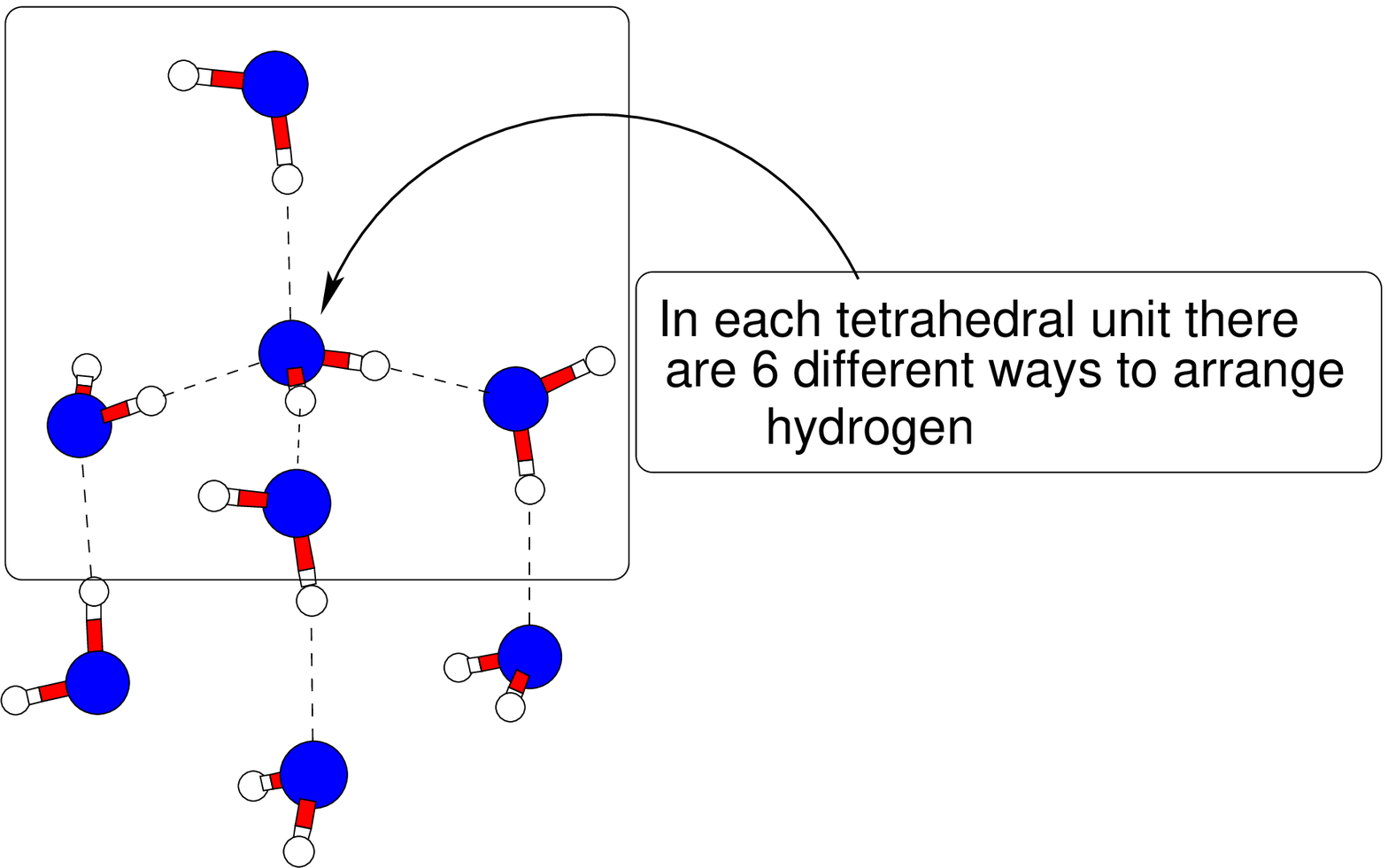}}
   \caption{ W.\ A.\ Adeagbo et. al.}
  \label{tetraice}
\end{figure}

\begin{figure} 
  \centering
  \resizebox{12cm}{!}{\includegraphics*{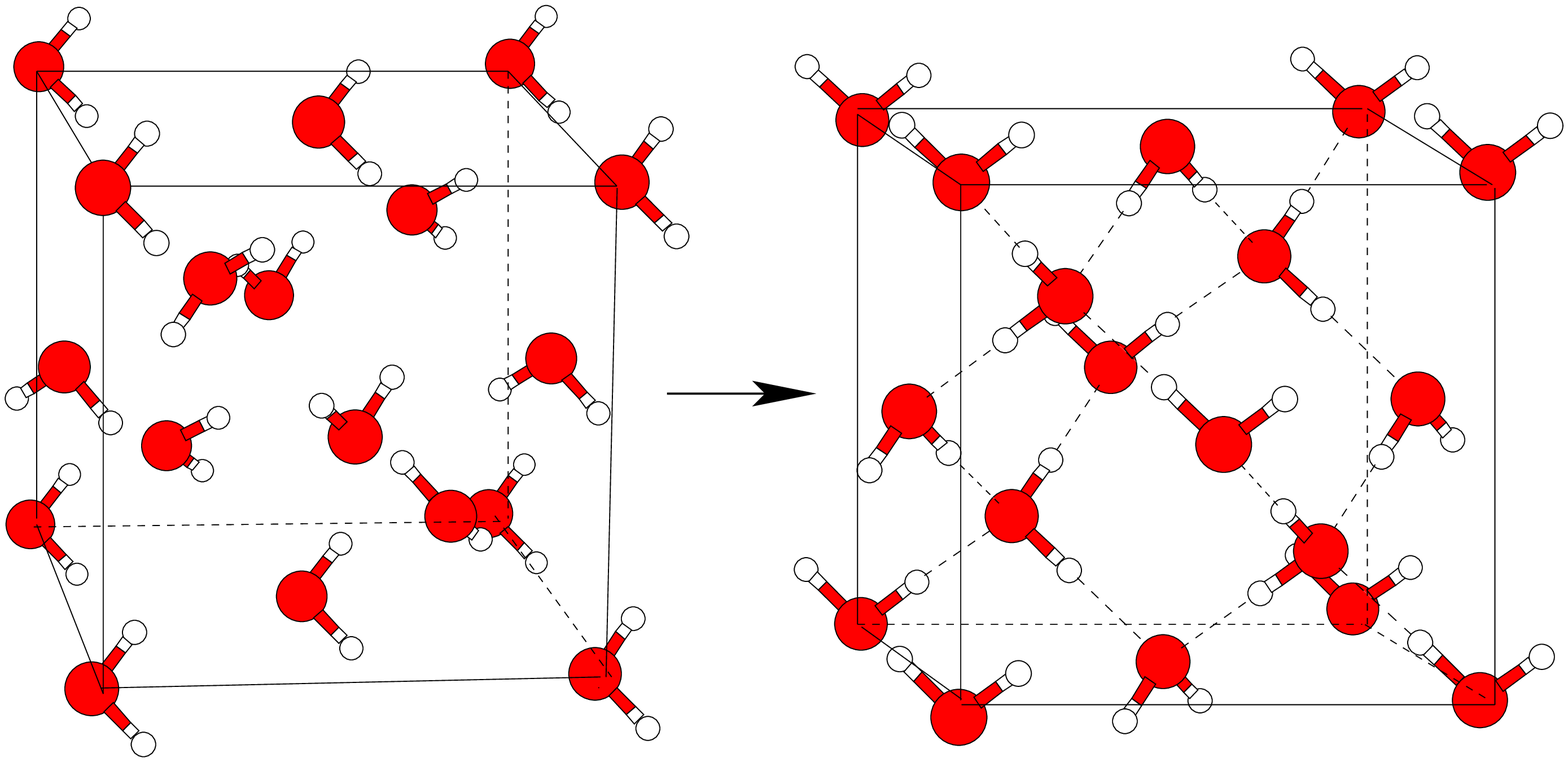}}
   \caption{W.\ A.\ Adeagbo et. al.}
\label{cellice}
\end{figure}

\begin{figure}
  \centering
  \resizebox{10cm}{!}{\includegraphics*{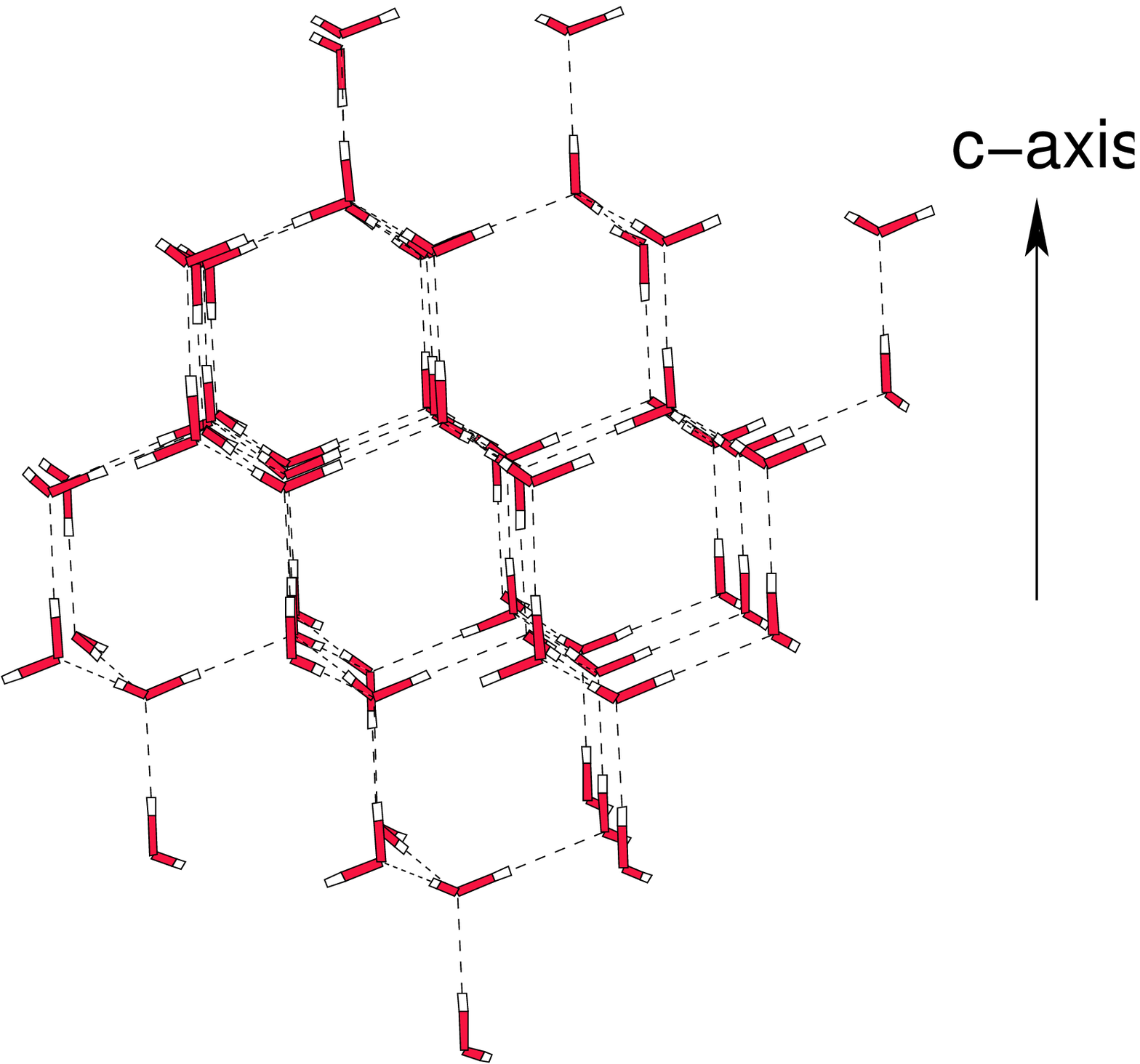}}
   \caption{W.\ A.\ Adeagbo et. al.}
  \label{ICE}
\end{figure}

\begin{figure} 
\hbox to \hsize{
\includegraphics*[height=6.0cm]{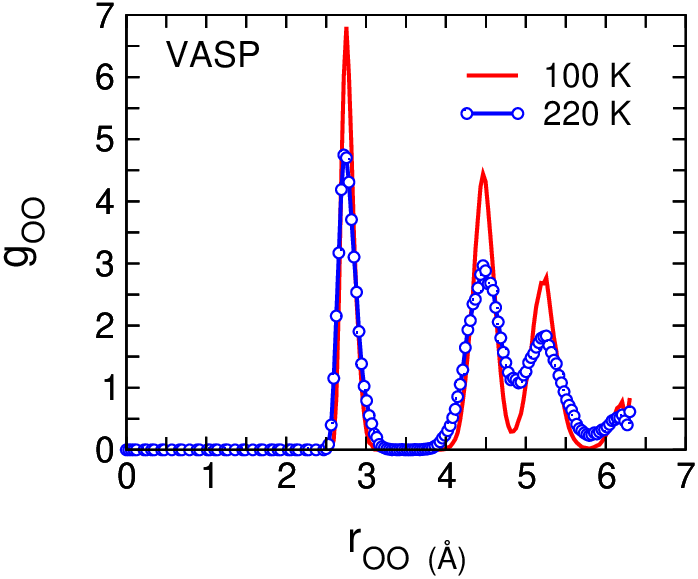}
\includegraphics*[height=6.0cm]{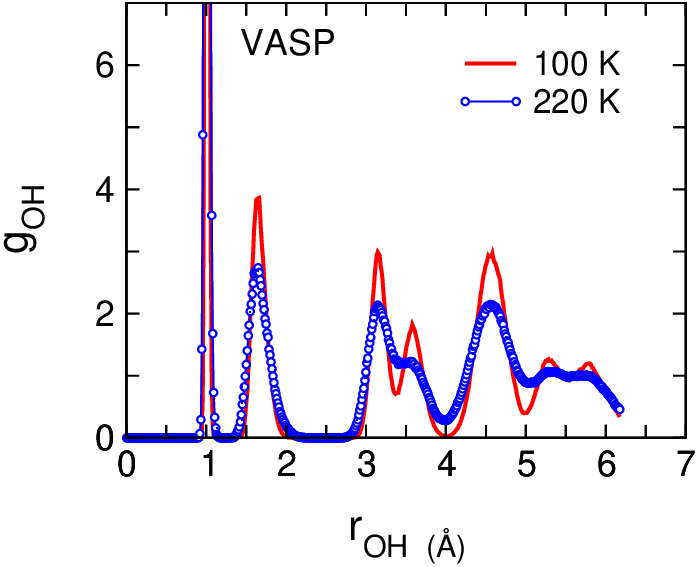}
}
  \caption {W.\ A.\ Adeagbo et. al.}
  \label{ICEGR100_OO_OH}
\end{figure}

\begin{figure}
  \centering
  \resizebox{9cm}{!}{\includegraphics*{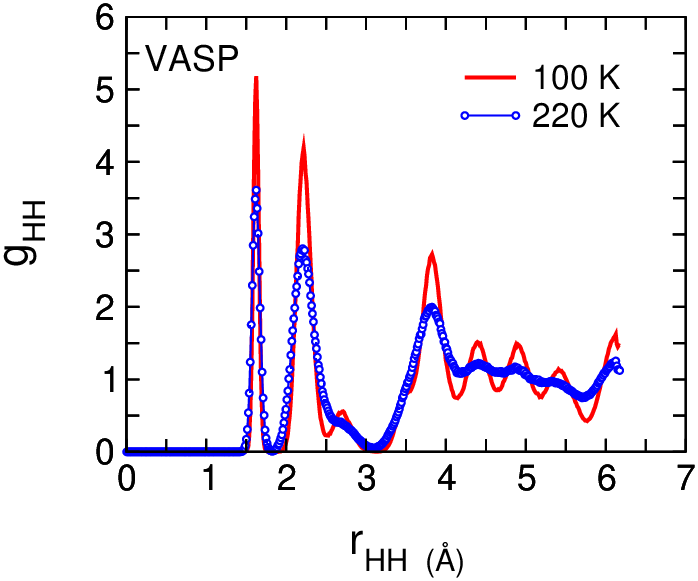}}
   \caption{W.\ A.\ Adeagbo et. al.}
  \label{ICEGR_100_HH}
\end{figure}

\begin{figure}
\hbox to \hsize{
\includegraphics*[height =5.4cm]{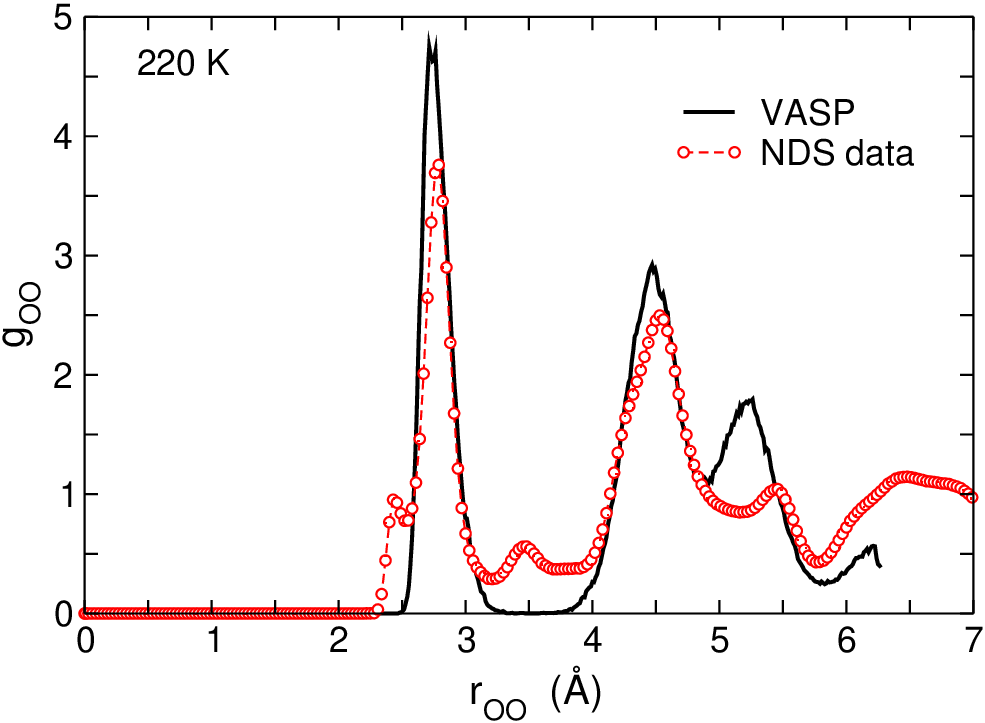}
\includegraphics*[height =5.4cm]{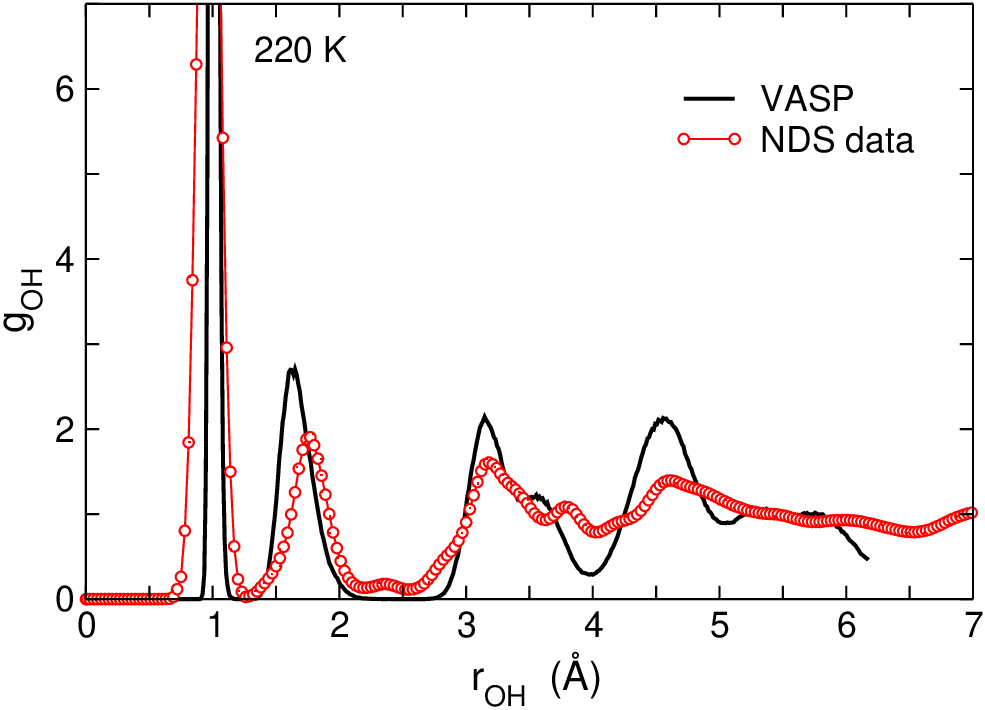}
}
  \caption
  {W.\ A.\ Adeagbo et. al.}
 \label{ICEGR_220_OO}
\end{figure}

\begin{figure}
 \centering
 \resizebox{13cm}{!}{\includegraphics*{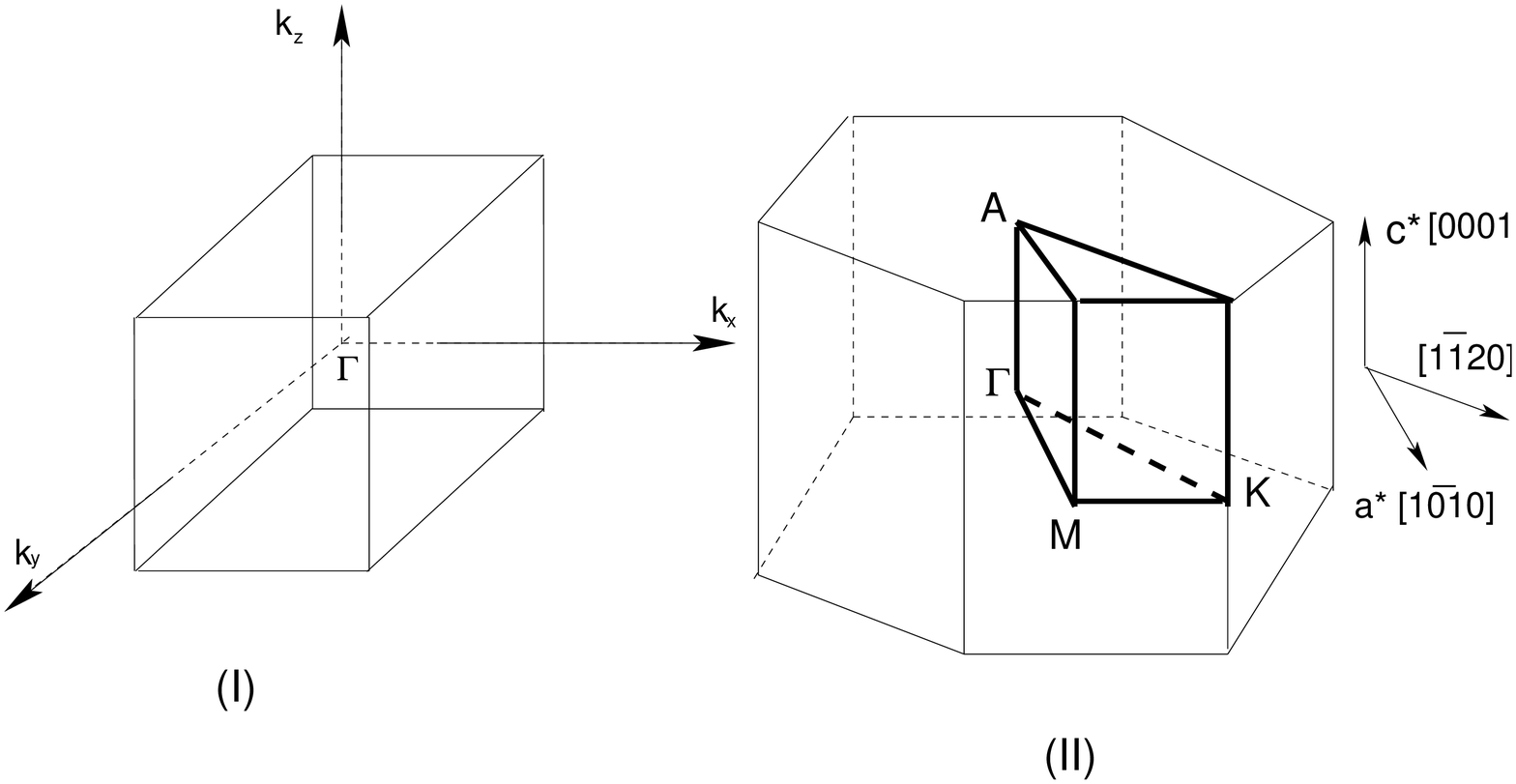}}
  \caption{W.\ A.\ Adeagbo et. al.}
 \label{hexsym}
\end{figure}

\begin{figure}[h]
 \centering
 \resizebox{13cm}{!}{\includegraphics*{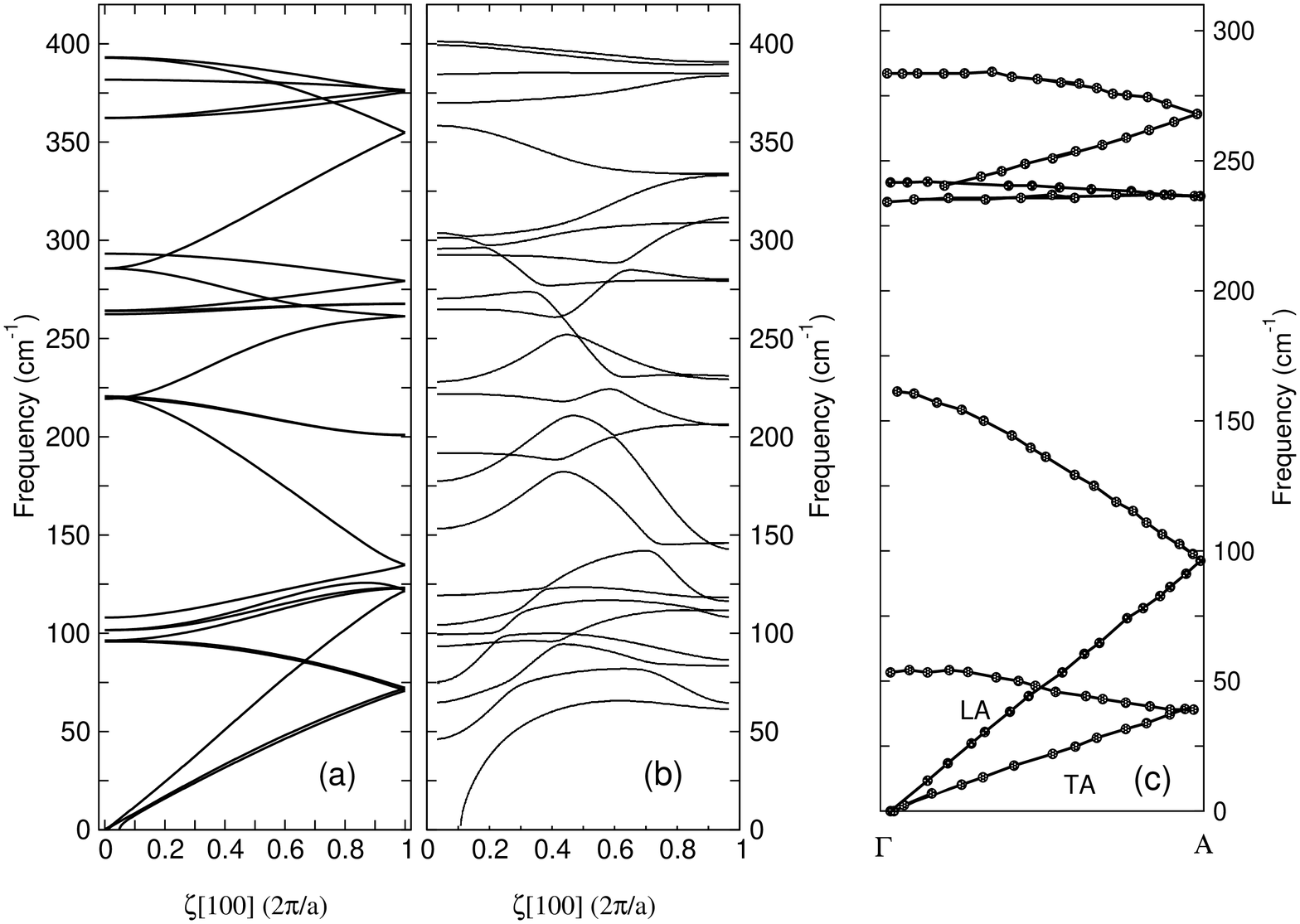}}
  \caption {W.\ A.\ Adeagbo et. al}
 \label{Cote}
\end{figure}

\begin{figure}
  \centering
  \resizebox{13cm}{!}{\includegraphics*{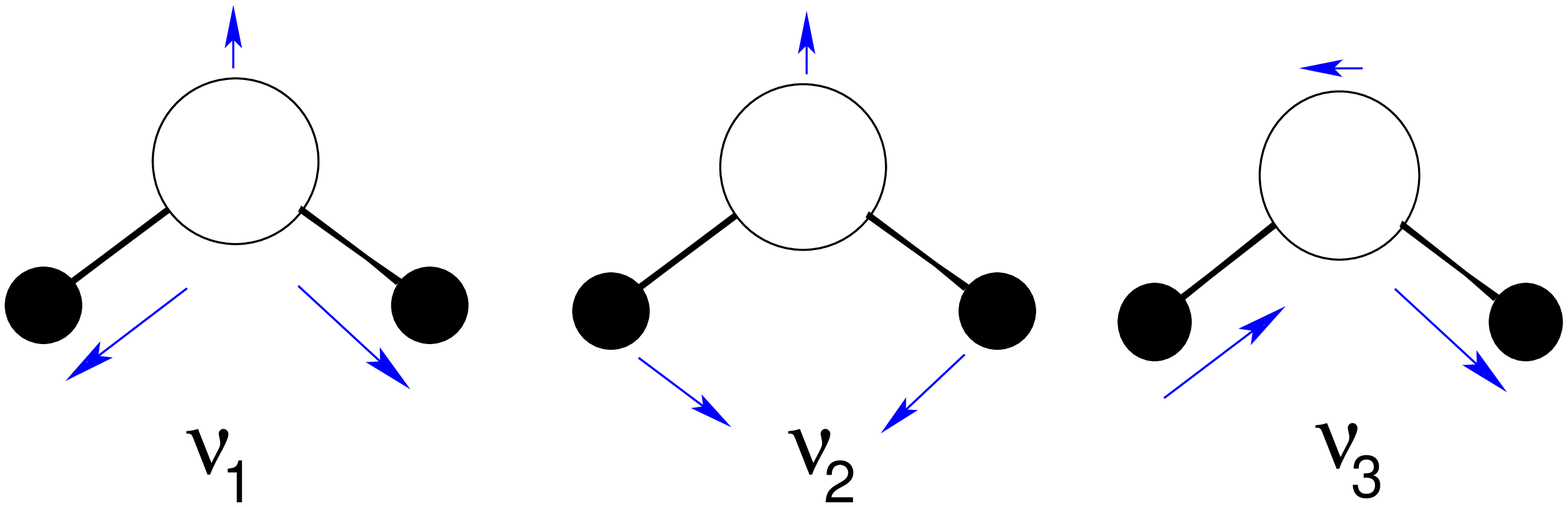}}
   \caption{W.\ A.\ Adeagbo et. al.}
  \label{tmodes}
\end{figure}

\begin{figure}
  \centering
  \resizebox{14cm}{!}{\includegraphics*{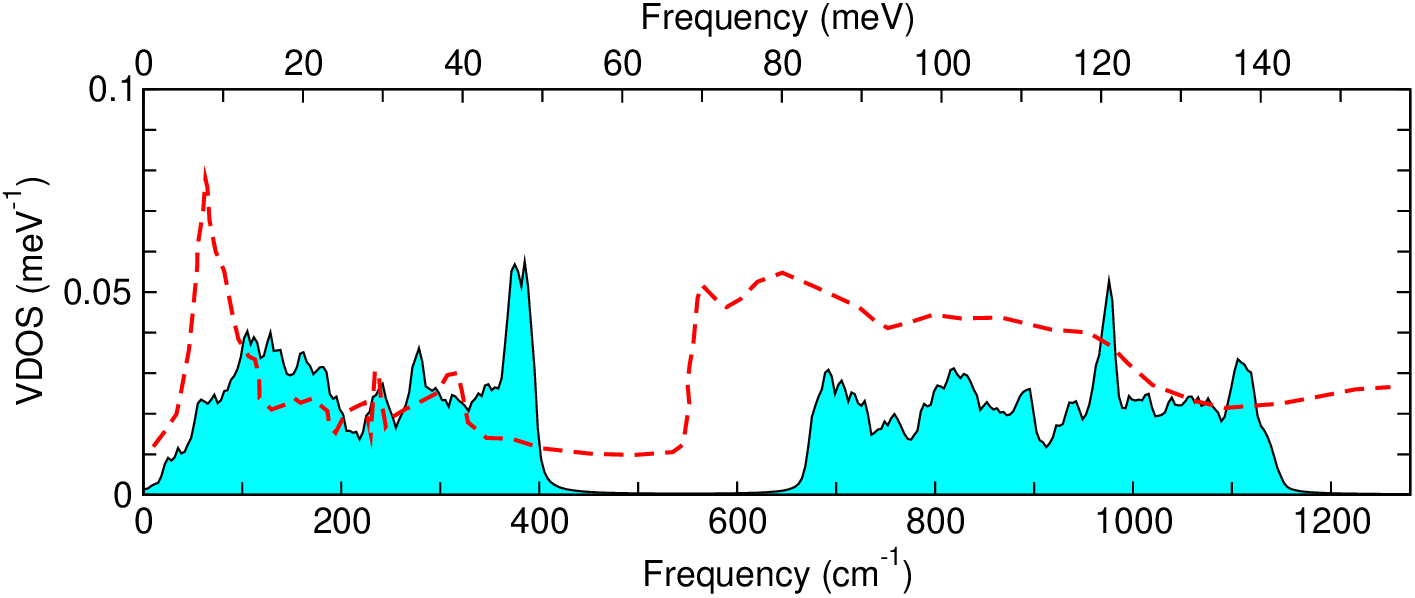}}
   \caption{W.\ A.\ Adeagbo et. al.}
 \label{totviblat2}
\end{figure}

\begin{figure}
  \centering
  \resizebox{11cm}{!}{\includegraphics*{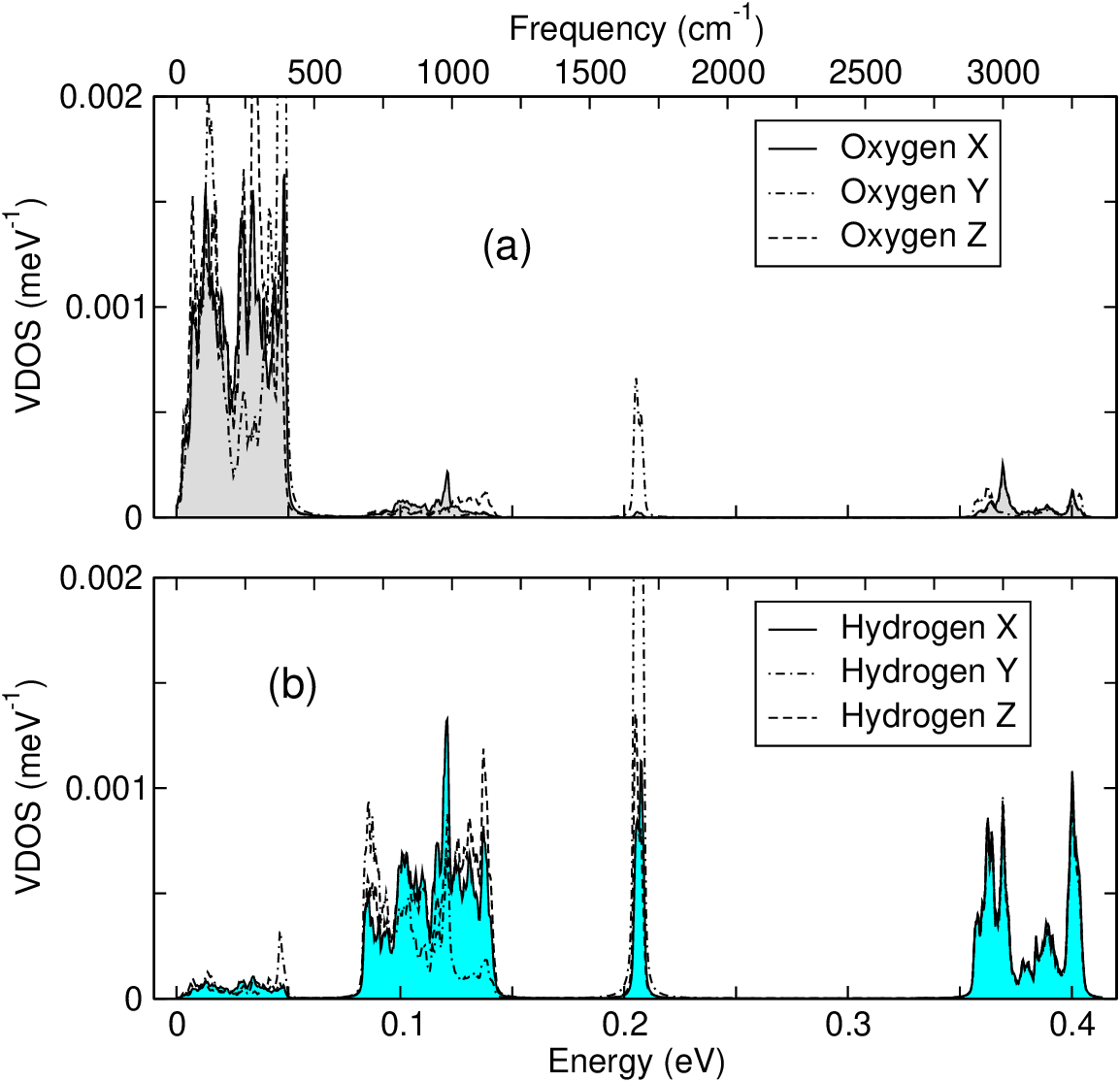}}
   \caption{W.\ A.\ Adeagbo et. al.}
  \label{viblat}
\end{figure}

\begin{figure}
  \centering
  \resizebox{11cm}{!}{\includegraphics*{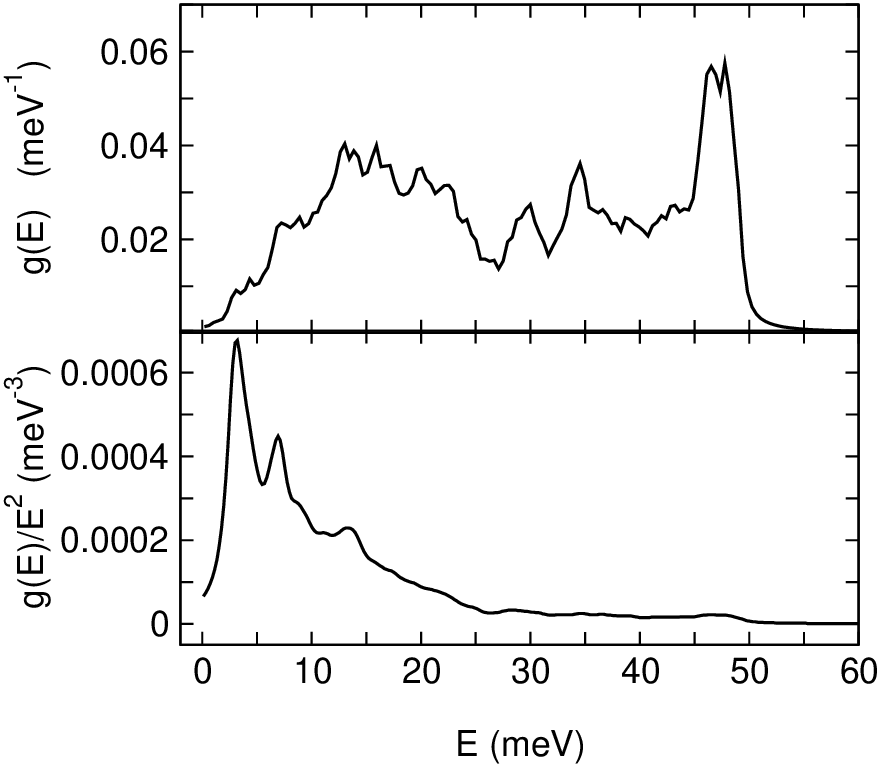}}
   \caption{W.\ A.\ Adeagbo et. al.}
  \label{Bosonp}
\end{figure}

\begin{list}{}{\leftmargin 2cm \labelwidth 1.5cm \labelsep 0.5cm}

\item[\bf Fig.~\ref{icephase}] The phase diagram of the stable phases of ice.

\item[\bf Fig.~\ref{tetraice}] The tetrahedral unit from which the
 hexagonal ice is created.

\item[\bf Fig.~\ref{cellice}]Initial and the relaxed geometry of the
  unit cell of ice. The ice structure was initially packed in a cubic
  unit cell with initial lattice constant taken from the
  literature~\citep*{cn:Lee1993} to be 6.35 {\AA}. There are no
  hydrogen bonds in the initial prepared structure shown on the left
  but were perfectly formed after the relaxation. The relaxed geometry
  has the values of $a \approx b \ne c$ which implies that the relaxed 
structure is tetragonal with ${c}/{a}$ ratio $\approx$ 0.988. 

\item[\bf Fig.~\ref{ICE}] Relaxed crystalline structure of ice
  produced from  the geometry in Fig.~\ref{tetraice} by replicating 
in all direction by the calculated lattice parameters.

\item[\bf Fig.~\ref{angdis}]Distribution of H-O-H angle in ice and
 water (in arbitrary units). The  comparison is done for the ice at
 220 K and liquid water simulated  at room temperature (298 K) and at
 high temperature in the supercritical regime. The H-O-H angles of ice
 structure are larger compared to those of  liquid water.

\item[\bf Fig.~\ref{ICEGR100_OO_OH}]Radial distribution functions
  $g_{\mathrm {OO}}$ and $g_{\mathrm {OH}}$ at 100 K and 220 K
  obtained with  VASP. There is a gradual loss of long-range order at 
 220 K as can be noticed in its 2nd and 3rd peaks of $\mathrm{g_{OO}}$
 when compared to the results for 100 K.

\item[\bf Fig.~\ref{ICEGR_100_HH}] Radial distribution functions
  $g_{\mathrm {HH}}$ at 100 K and 200 K obtained with VASP.  There is
  a gradual ``loss of peaks'' observed for temperature 100 K at 2.5,
  4.4, 4.8 and 5.4 {\AA} due to the loss of long-range order as the 
 temperature increases. 

\item[\bf Fig.~\ref{ICEGR_220_OO}]Radial distribution functions
  $g_{\mathrm {OO}}$ and $g_{\mathrm {OH}}$ at 220 K obtained  with
  VASP compared to neutron diffraction scattering data (NDS)
  ~\citep*{cn:Soper2000}. 

\item[\bf Fig.~\ref{ICEGR_220_HH}]Radial distribution functions
  $g_{\mathrm {HH}}$  obtained with VASP at 220 K compared to neutron
  diffraction scattering data (NDS) ~\citep{cn:Soper2000}.

 \item[\bf Fig.~\ref{hexsym}] (I) Brillouin zone for the cubic
 symmetry used in our VASP calculation of the ice  system. Our phonon
 dispersions are calculated in [100] $k_x$-direction, (II) The first
 Brillouin  zone for the  structure of
 ice Ih with origin at the point $\Gamma$. $\Gamma$A =$\frac{1}{2}c^*$
 and  $\Gamma$M =$\frac{1}{2}a^*$, where $a^*$ and $c^*$ are the
 vectors of the  reciprocal lattice ~\citep{cn:Victor1999}. The dispersion
 curves are commonly drawn along the lines of symmetry $\Gamma$A, $\Gamma$M,
 and $\Gamma$K. 

\item[\bf Fig.~\ref{Cote}](a) With VASP calculated phonon dispersion
  curves of ice in [100] direction of the tetragonal unit cell
  compared to (b) dispersion relations in the translational frequency
  range for the ice structure plotted along the Cartesian directions 
from zone center to zone edge in the Brillouin zone of the orthorhombic
eight-molecule unit cell (Cote {\it et. al.}~\citep*{cn:Cote2003}) and
  (c) the experimental dispersion of D$_2$O ice according to  Renker's
  model ~\citep*{cn:Renker1973}. The difference in scale of (c) from
  (a) and (b) is due to the isotopic  effect because of difference in
  the masses  of  hydrogen and Deuterium atoms. 

\item[\bf Fig.~\ref{tmodes}]The three normal modes of an isolated
  water molecule. Motion with frequency $\nu_1$ can be regarded as
  symmetric stretch, $\nu_2$ as  bending and $\nu_3$ as
  anti-symmetric~\citep*{cn:Victor1999}. 

\item[\bf Fig.~\ref{totviblat}]Total vibrational density of states
  (VDOS) of ice  based on the lattice dynamics together with full
  phonon dispersion curves.   The VDOS show all the important regions
  such as the intermolecular translational, librational, bending and
  the stretching frequency range.

\item[\bf Fig.~\ref{totviblat2}]Enlargement of calculated total
  vibrational density of states in Fig.~\ref{totviblat} showing the
  intermolecular range of frequencies. The broken line is taken from
  the inelastic neutron scattering data available through
  Ref.~\citep*{cn:Cote2003}.

\item[\bf Fig.~\ref{viblat}] Partial density of states of ice based on
  the lattice dynamics. (a) shows the $x$, $y$ and $z$ components of
  VDOS for the oxygen atoms in the  H$_2$O ice.
(b) shows the $x$, $y$ and $z$  VDOS for corresponding hydrogen atoms for the
whole range of frequencies. It is interesting to notice that in the intra
molecular bending region (1580-1680 cm$^{-1}$), only one of the  components of
the VDOS of hydrogen (say $y$) dominates while $x$ and $z$ contribute less. The
contribution from the oxygen atoms can be regarded as being completely
recessive in this region.

\item[\bf Fig.~\ref{Bosonp}] Plot of the VDOS $g(E)$  and the
  corresponding $g(E)/E^2$ vs. $E$ for the  region of
 translational mode.  The boson peak  is found in the low-energy
 region at 3.5 meV. 

 \end{list}


\clearpage

\begin{table} 
\begin{minipage}{\textwidth}
\centering
\caption{ W.\ A.\ Adeagbo et. al.}
\vspace{0.4cm}
\begin{tabular*}{0.75\textwidth}{@{\extracolsep{\fill}} l c c c c c c}
\hline
\hline
$\mathrm{\times 10^3 \ m/s}$ && Experiment\footnote{~\citep*{cn:Gammon1983}} && Theory\\
  \hline
$v_\mathrm{LA}$ && 4.04 && 4.86 & \\
$v_\mathrm{TA}$ && 1.80 && 3.02 & \\ \hline
\end{tabular*}
\label{velTab}
\end{minipage}
\end{table}

\clearpage

 \begin{figure}
  \centering
  \resizebox{8cm}{!}{\includegraphics*{adeagbo-fig-5.eps}}
   \caption{W.\ A.\ Adeagbo et. al.}
  \label{angdis}
\end{figure}

\begin{figure}
  \centering
  \resizebox{10cm}{!}{\includegraphics*{adeagbo-fig-9.eps}}
   \caption{W.\  A.\ Adeagbo et. al.}
 \label{ICEGR_220_HH}
\end{figure}

\begin{figure}
  \centering
  \resizebox{14cm}{!}{\includegraphics*{adeagbo-fig-13.eps}}
   \caption{W.\ A.\ Adeagbo et. al.}
  \label{totviblat}
\end{figure}

\end{document}